\documentclass[aip,amsmath,amssymb,preprint]{revtex4-1}
\usepackage{graphicx}
\draft 

\begin{document}


\title{\textit{Ab-initio} quantum transport simulation of self-heating
  in single-layer 2-D materials} 



\author{Christian Stieger, Aron Szabo, Teut\"e Bunjaku, and Mathieu Luisier}
\email[]{mluisier@iis.ee.ethz.ch}
\affiliation{Integrated Systems Laboratory, ETH Z\"urich, Gloriastrasse 35, 8092 Z\"urich, Switzerland}


\date{\today}

\begin{abstract}
Through advanced quantum mechanical simulations combining electron and
phonon transport from first-principles self-heating effects are
investigated in $n$-type transistors with a single-layer MoS$_2$, WS$_2$, and
black phosphorus as channel materials. The selected 2-D crystals all
exhibit different phonon-limited mobility values, as well as electron
and phonon properties, which has a direct influence on the increase of
their lattice temperature and on the power dissipated inside their
channel as a function of the applied gate voltage and electrical
current magnitude. This computational study reveals (i)
that self-heating plays a much more important role in 2-D materials
than in Si nanowires, (ii) that it could severely limit the
performance of 2-D devices at high current densities, and (iii) that  
black phosphorus appears less sensitive to this phenomenon than
transition metal dichalcogenides.
\end{abstract}


\maketitle 

\section{Introduction}

Only a fraction of the existing 2-D materials has been identified so
far \cite{mounet}, probably less than 50 components have been
mechanically exfoliated or grown via chemical vapor deposition (CVD),
but it already clearly appears that the 2-D technology might have a
positive impact in various domains of application, one of them being the
search for next-generation logic switches beyond Si FinFETs
\cite{kuhn}. While graphene, the first experimentally demonstrated 2-D
compound \cite{novoselov}, is not suitable as transistor channel due
to the absence of a bandgap, monolayers of MoS$_2$ and other
transition metal dichalcogenides (TMDs) look very promising as silicon
replacement in upcoming ultra-scaled transistors \cite{kis}: they
exhibit excellent electrostatic properties due to their 2-D nature,
relatively large band gaps ($E_g\sim$1.8 eV in MoS$_2$), high
$I_{ON}$/$I_{OFF}$ current ratios, and decent room temperature
single-layer mobility values, e.g. 214 cm$^2$/Vs for
WS$_2$ encapsulated between two h-BN layers \cite{iqbal}. Furthermore,
because the atomic composition and crystallographic phase of TMDs
fully determine whether they behave as semiconductors, insulators, or 
metals \cite{miro}, active components relying solely on stacks of such
materials can be easily imagined for the future.

The ``current vs. voltage'' characteristics and electron/hole
mobilities of devices based on single-layer TMDs and other 2-D
materials have been extensively investigated in the recent past,
contrary to their thermal and electro-thermal properties
(self-heating, the formation of local hot spots, breakdown failures),
which have remained largely unexplored up to now. It is however
expected that the influence of thermally induced effects becomes
significant in ultra-scaled structures due to the close proximity of
electrons and phonons and their potentially increased coupling
\cite{pop0}. Few studies, mainly concerned with graphene
\cite{balandin,shi,pop1}, but also molybdenum disulfide
\cite{xing,pop2} and black phosphorus \cite{cahill} have discussed
these issues. Still, it can be generally said that the thermal
behavior of 2-D materials is not completely understood yet and that
rapid progresses, both at the theoretical and experimental levels, are
required to shed light on the underlying physics and to enable the
emergence of 2-D devices with improved electro-thermal
functionalities, e.g. nanoscale thermoelectric generators or Peltier 
coolers.

This paper intends to give a theoretical insight into the
electro-thermal properties of transistors with a single-layer 2-D
material as channel. For that purpose, advanced simulation techniques
will be utilized. On the electrical side, the state-of-the-art
consists in utilizing ballistic full-band solvers implementing the
Non-equilibrium Green's Function formalism (NEGF) \cite{datta}. The
necessary Hamiltonian matrix relies either on semi-empirical
(tight-binding) \cite{hguo,klimeck} models, it is derived
from first-principles via maximally localized Wannier functions
\cite{register,fiori1}, or it is constructed at the \textit{ab-initio}
level \cite{qw0}. With many TMDs, such approaches induce non-physical
negative differential resistance (NDR) behaviors that have never been
experimentally observed at room temperature \cite{register}. The NDR
originates from the TMD bandstructure, which exhibits several narrow
energy bands that cannot carry any current if the electrostatic potential
undergoes large variations between the source and drain contacts
\cite{gnani}. It is therefore an artifact of the ballistic
approximation. The inclusion of electron-phonon scattering helps get
rid of NDR by connecting bands that would otherwise remain independent
from each other \cite{szabo1}. Combining a full-band approach with
electron-phonon scattering has been applied in several occasions to
2-D materials, but only to extract their electron mobility
\cite{kaas,qw,marzari,fiori2}, not to obtain their current
characteristics as a function of the externally applied voltages.

On the thermal modeling side, it is widely recognized that the tiny
dimensions of 2-D materials require a quantum mechanical approach to
treat heat transport, which should be done at the phonon level. Two
different methods have been successfully applied by various research
groups to obtain the thermal conductivity of single-layer 2-D
materials: (i) molecular dynamics (MD) simulations \cite{md1,md2} and
(ii) the coupling of density-functional theory (DFT) with the linearized
Boltzmann Transport equation (LBTE), where the former provides the
bandstructure and scattering parameters for the latter
\cite{lbt1,lbt2,marzari2}. The advantage of MD over DFT+LBTE is that
it allows for the consideration of non-homogeneous atomic
structures with spatial or material variations. Nevertheless, neither
one nor the other model lends itself naturally to a self-consistent
coupling with electron transport. The NEGF formalism offers more
flexibility for that because it can capture the energy
exchanges between the electron and phonon populations via scattering
self-energies, thus automatically accounting for local lattice
temperature increases \cite{asai,pecchia,frederiksen,jtlu1,jtlu2}. 

Here, by combining the electro-thermal simulation framework of
Ref.~\cite{prb14}, originally developed for nanowire transistors, and
the \textit{ab-initio} modeling platform of Ref.~\cite{szabo1},
specifically dedicated to the investigation of 2-D materials, the
first theoretical results of electro-thermal transport through logic
devices with a 2-D monolayer semiconductor as channel will be
demonstrated. To the best of our knowledge such an accurate technique
had not been tested in the context of 2-D materials before, mainly
because its application is normally limited to small systems composed
of 100 atoms or less due to the high computational burden associated
with the description from first-principles of electron and phonon
bandstructures. The paper is organized as follows: after a
detailed presentation of the modeling
approach in Section \ref{sec:method}, simulation results will be
introduced in Section \ref{sec:results} for transistors with a
single-layer MoS$_2$, WS$_2$, and black phosphorus channel. The
low-field phonon-limited mobility, \textit{I-V} characteristics,
lattice temperature, and dissipated power of these devices will be
analyzed and compared to each other. Finally, conclusions will be
drawn in Section \ref{sec:conc}. 

\section{Simulation Approach}\label{sec:method}

\subsection{Electron and Phonon Transport}

\begin{figure}
\includegraphics[width=0.45\linewidth]{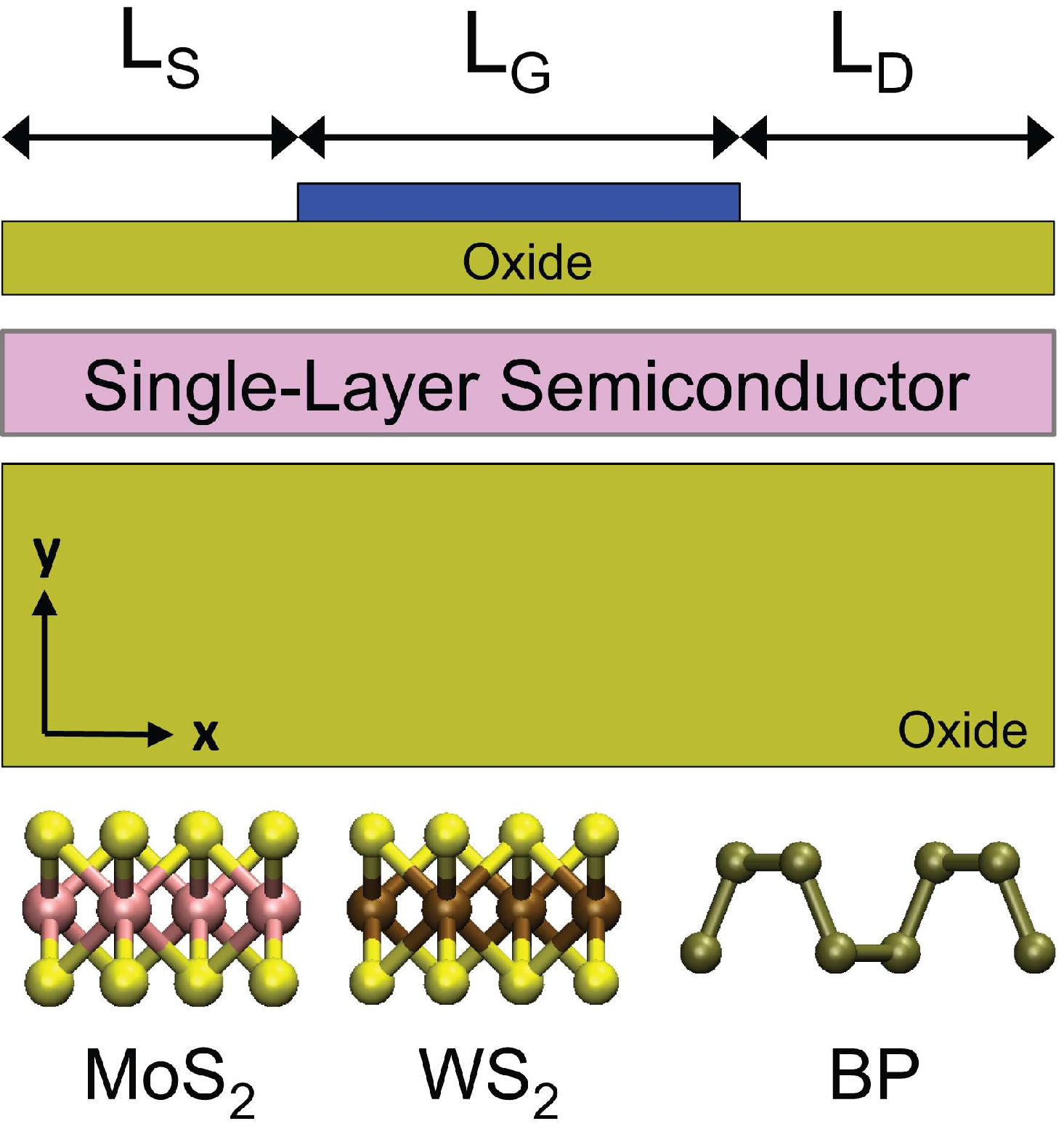}
\caption{\label{fig:fig1} (Color online) Schematic view of the
  single-gate field effect transistor with single-layer MoS$_2$,
  WS$_2$, or armchair-oriented black phosphorus (AC BP) as semiconducting
  channel. The gate length is set to $L_g$=15 nm, while the source and
  drain extensions measure $L_s$=$L_d$=12.5 nm with a donor doping
  concentration $N_D$=6e13 cm$^{-2}$. Perfectly ohmic contacts are
  considered. The gate contact is separated from the channel by an
  HfO$_2$ dielectric layer of thickness $t_{ox}$=3 nm and permittivity
  $\epsilon_R$=20. Transport occurs along the $x$-axis, $y$ is the
  direction of confinement, and the $z$-axis (out-of-plane) is assumed
  periodic and modeled via a $k_z$/$q_z$ momentum dependence of the physical
  quantities.}
\end{figure}

Transistor structures similar to the one depicted in
Fig.~\ref{fig:fig1} have been simulated in this work. Transport occurs
along the $x$-axis, $y$ is a direction of confinement, and the
out-of-plane dimension, $z$, is assumed periodic and gives rise to a
$k_z$(electron momentum along $z$)- or $q_z$(phonon momentum along
$z$)-dependence of all computed physical quantities. The electron and
phonon properties are modeled at the \textit{ab-initio} level with
NEGF. Their coupling is realized via scattering self-energies that
ensure current and energy conservation. In this context, the following
system of equations must be solved for the electron population
\begin{eqnarray}
\left\{
\begin{array}{l}
\sum_l\left(\mathbf{E}\delta_{li}-\mathbf{H}_{il}(k_z)-\mathbf{\Sigma}^{RB}_{il}(E,k_z)-
\mathbf{\Sigma}^{RS}_{il}(E,k_z)\right)\cdot \mathbf{G}^{R}_{lj}(E,k_z)=\delta_{ij}\\
\mathbf{G}^{\gtrless}_{ij}(E,k_z)=\sum_{lm}\mathbf{G}^{R}_{il}(E,k_z)\cdot
\left(\mathbf{\Sigma}^{\gtrless B}_{lm}(E,k_z)+\mathbf{\Sigma}^{\gtrless S}_{lm}(E,k_z) 
\right)\cdot \mathbf{G}^{A}_{mj}(E,k_z).
\end{array}
\right.
\label{eq:1}
\end{eqnarray}
In Eq.~(\ref{eq:1}), $\mathbf{E}$ is a diagonal matrix that contains
the electron energy $E$ as entry. The $\mathbf{G}_{ij}(E,k_z)$'s represent the
electron Green's Functions at energy $E$ and momentum $k_z$ between
atoms $i$ and $j$ situated at position $\mathbf{R}_i$ and $\mathbf{R}_j$,
respectively. They are of size $N_{orb,i}\times N_{orb,j}$, where
$N_{orb,i}$ is the number of orbitals (basis components) describing
atom $i$. The $\mathbf{G}_{ij}(E,k_z)$'s can be either retarded ($R$),
advanced ($A$), lesser ($<$), or greater ($>$). The same conventions
apply to the self-energies $\mathbf{\Sigma}_{ij}(E,k_z)$ whose
additional index $B$ ($S$) stands for boundary (scattering). Details
about the  $\mathbf{\Sigma}_{ij}(E,k_z)$'s are provided below.

\begin{figure}
\includegraphics[width=\linewidth]{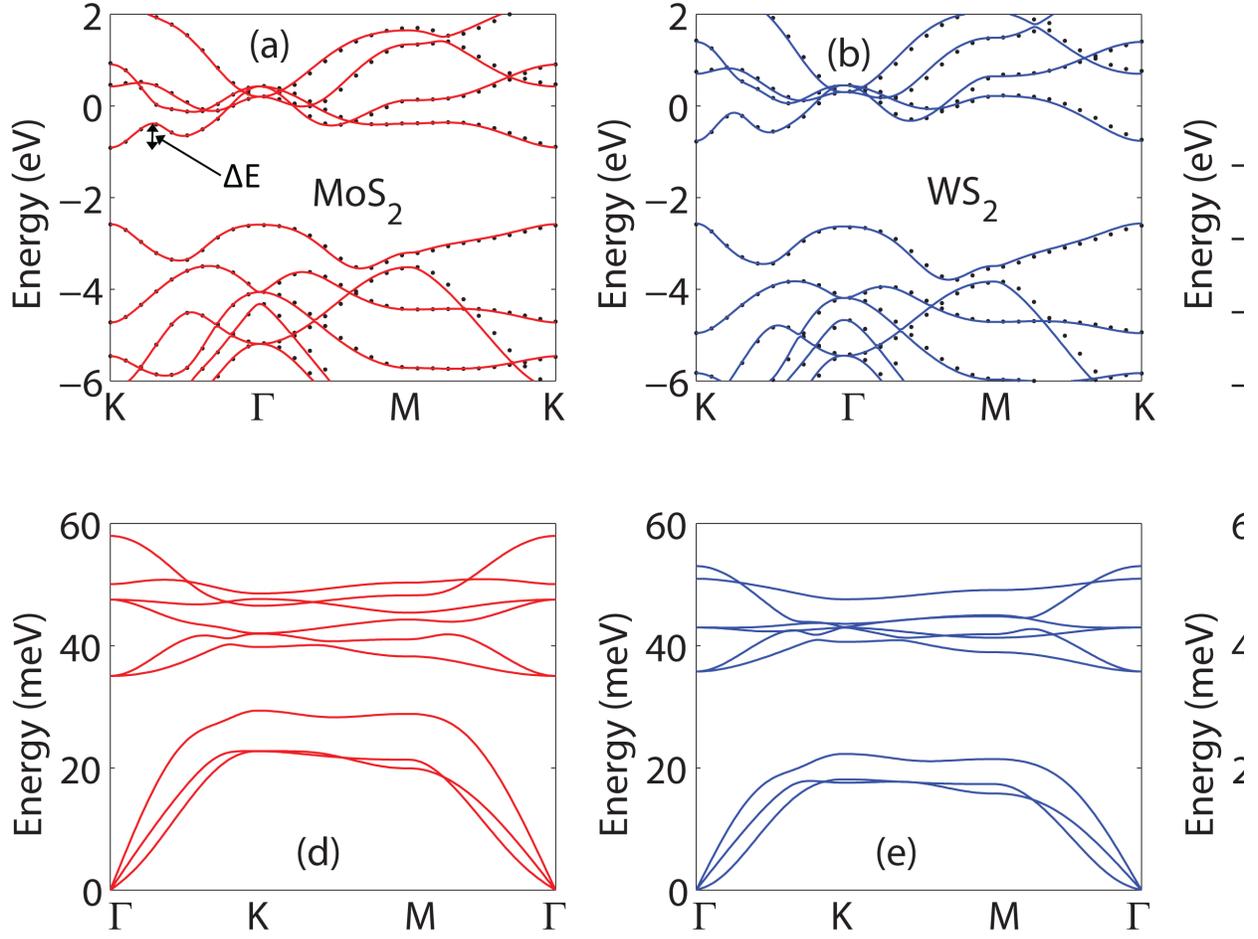}
\caption{\label{fig:fig2} (Color online) Electron (a-c) and phonon
  (d-f) bandstructure of single-layer MoS$_2$ (a and d), WS$_2$ (b and
  e), and black phosphorus (c and f) along their high symmetry
  lines. In case of electrons, the black dots are the results of DFT
  calculations with VASP \cite{vasp}, whereas the solid lines
  represent the bandstructures after a conversion of the plane-wave
  outputs of VASP into a set of maximally localized Wannier functions
  with the wannier90 tool \cite{wannier90}. The phonon modes have been
  computed with DFPT \cite{dfpt}.}
\end{figure}

The Hamiltonian matrix elements $\mathbf{H}_{il}(k_z)$ are key
ingredients in Eq.~(\ref{eq:1}). To simulate 2-D materials, they are
usually either expressed in a semi-empirical basis such as
tight-binding \cite{hguo,klimeck} or in a localized
density-functional theory (DFT) basis \cite{qw0}. Here, an
intermediate scheme is used: after identifying a small unit
cell that is representative for the considered 2-D crystal, a DFT
calculation is performed with a standard tool such as VASP
\cite{vasp}. The plane-wave outputs of VASP are then converted into a
set of maximally localized Wannier functions (MLWFs) with the
wannier90 package \cite{wannier90}. This allows for the creation of a
tight-binding-like Hamiltonian matrix that is scaled up to
match the desired structure dimensions. Finally, quantum transport 
simulations are run with the produced Hamiltonian as input. The whole
process is highlighted in Ref.~\cite{szabo1}. Note that MoS$_2$ and
WS$_2$ have been treated here within the generalized gradient
approximation (GGA) of Perdew, Burke, and Ernzerhof (PBE) \cite{pbe},
whereas the hybrid functional of Heyd, Scuseria, and Ernzerhof (HSE06)
\cite{hse06} has been employed for black phosphorus. A
25$\times$1$\times$25 Monkhorst-Pack k-point grid and a 500 eV
plane-wave cutoff energy have been enforced in all electronic
structure calculations. Spin-orbit coupling has been 
neglected. The accuracy of the plane-wave $\rightarrow$ MLWF
conversion is demonstrated in Fig.~\ref{fig:fig2}(a-c) for the
selected 2-D materials: 3 $p$-like orbitals have been adopted for S, 5
$d$-like orbitals for Mo and W, and 4 $sp^3$ hybrid for P.

To handle phonon transport through transistors with a 2-D channel, as
in Fig.~\ref{fig:fig1}, the following NEGF-based system of equations
must be processed 
\begin{eqnarray}
\left\{
\begin{array}{l}
\sum_l\left(\mathbf{\omega}^2\delta_{li}-\mathbf{\Phi}_{il}(q_z)- 
\mathbf{\Pi}^{RB}_{il}(\omega,q_z)-\mathbf{\Pi}^{RS}_{il}(\omega,q_z)\right)
\cdot \mathbf{D}^{R}_{lj}(\omega,q_zz)=\delta_{ij}\\
\mathbf{D}^{\gtrless}_{ij}(\omega,q_z)=\sum_{lm}\mathbf{D}^{R}_{il}(\omega,q_z)
\cdot\left(\mathbf{\Pi}^{\gtrless B}_{lm}(\omega,q_z)+\mathbf{\Pi}^{\gtrless S}_{lm}(\omega,q_z)
\right)\cdot \mathbf{D}^{A}_{mj}(\omega,q_z),
\end{array}
\right.
\label{eq:2}
\end{eqnarray}
where the $\mathbf{D}(\omega,q_z)$'s are the phonon Green's functions
at frequency $\omega$ and momentum $q_z$, $\mathbf{\omega}$ a diagonal
matrix containing the phonon frequency as single entry, the
$\mathbf{\Pi}^{B}(\omega,q_z)$'s the boundary self-energies, and the
$\mathbf{\Pi}^{S}(\omega,q_z)$'s the scattering ones, while
$\mathbf{\Phi}(q_z)$ refers to the dynamical (Hessian) matrix of the
studied domain. The same Green's Function types as for electrons also
exist for phonons (retarded, advanced, lesser, and greater). All blocks
involved in Eq.~(\ref{eq:2}) are of size 3$\times$3, which corresponds
to the degrees of freedom of the crystal vibrations. The matrix
$\mathbf{\Phi}(q_z)$ is constructed in an analogous way as
$\mathbf{H}(kz)$: through density functional perturbation theory
(DFPT) \cite{dfpt}, as available in VASP, the dynamical matrix of a
representative unit cell is computed from first-principles. It is then
scaled up to obtain the entries for a larger device structure,
e.g. the transistor in Fig.~\ref{fig:fig1}. The phonon bandstructures
of single-layer MoS$_2$, WS$_2$, and black phosphorus are displayed in
Fig.~\ref{fig:fig2}(d-f).

\begin{figure}
\includegraphics[width=0.45\linewidth]{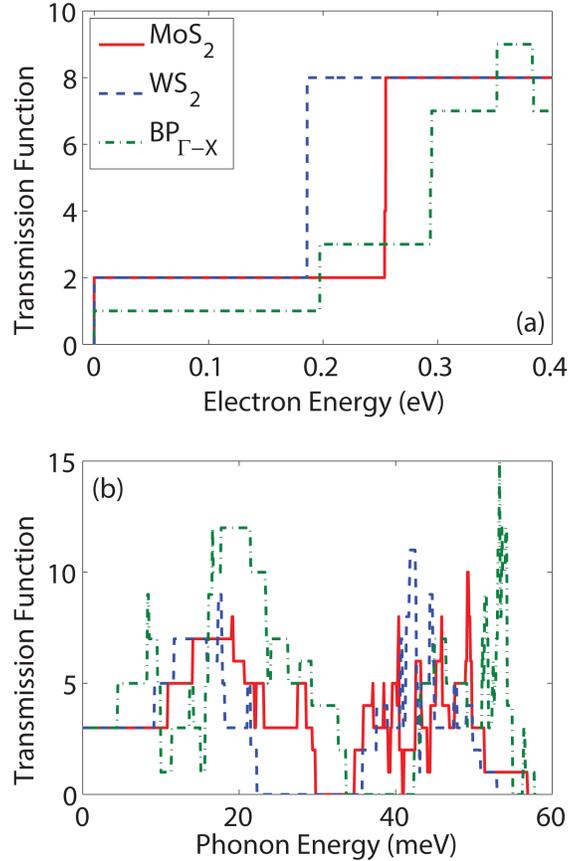}
\caption{\label{fig:fig3} (Color online) (a) Electron transmission
  function, as obtained from quantum transport calculations, through
  single-layer MoS$_2$ (solid red line), WS$_2$ (dashed blue line),
  and AC BP (dashed-dotted green line) at $k_z$=0, with flat-band
  conditions, and with the conduction band minimum aligned with the
  energy $E$=0 eV. (b) Same as (a), but for phonons at $q_z$=0.}
\end{figure}

With the knowledge of $\mathbf{H}(k_z)$ and $\mathbf{\Phi}(q_z)$,
electron and phonon quantum transport simulations can be performed,
first in the ballistic limit of transport,
i.e. $\mathbf{\Sigma}^{S}(E,k_z)$=$\mathbf{\Pi}^{S}(\omega,q_z)$=0. Eqs.~(\ref{eq:1})
and (\ref{eq:2}) can be solved with a recursive Green's Function
(RGF) algorithm \cite{rgf} that produces only the desired entries of
the Green's Functions. The open boundary conditions are derived from
the solution of eigenvalue problems with a shift-and-invert technique
\cite{prb06}. As an illustration, the ballistic electron and phonon transmission
functions through 40 nm long monolayers of MoS$_2$, WS$_2$, and black
phosphorus (in the armchair configuration) have been computed and the
results are plotted in Fig.~\ref{fig:fig3} (a) and (b) for $k_z$=0 and
$q_z$=0, respectively. A flat band potential is assumed in all cases
so that a typical step-like behavior of the transmission as a
function of the electron/phonon energy can be observed. Attention
should be paid to the fact that the mini-gaps (regions with no bands)
present in the phonon bandstructures in Fig.~\ref{fig:fig2}(d-f) are
clearly visible in the transmission functions in Fig.~\ref{fig:fig3}(b).

To accurately compute the current that flows through a 2-D transistor,
it is not sufficient to keep one single momentum point,
e.g. $k_z$=$q_z$=0. Here, it has been found that a total of 11 $k_z$
and $q_z$ momentum points is necessary to reliably model the
periodicity of the out-of-plane direction $z$ (no more than 1\%
current variations as compared to simulations with 21 momentum
points). It should finally be underlined that electrons and
phonons can only enter and leave the simulation domain at the source
and drain contact extremities, no escape through the top or bottom
oxide of the transistor in Fig.~\ref{fig:fig1} is possible. This
restriction certainly causes an overestimation of the lattice
temperature \cite{apl17} when the electron and phonon populations
fully interact with each other, as described in the next Section.

\subsection{Electron and Phonon Coupling}

To complete the picture it remains to define the scattering
self-energies that couple electron and phonon transport and drive both
populations out-of-equilibrium. To reduce the
computational intensity, only the diagonal blocks of the
electron-phonon components $\mathbf{\Sigma}^{\gtrless S}(E,k_z)$ are
retained. It can be analytically demonstrated, as in Ref.~\cite{jtlu1},
that these scattering self-energies ensure total energy
conservation. Each individual block is defined as
\begin{eqnarray}
\mathbf{\Sigma}^{\gtrless S}_{nn}(E,k_z)&=&i\sum_{q_z}\sum_{l}\sum_{ij}
\int\frac{d(\hbar\omega)}{2\pi}\left[\nabla_i\mathbf{H}_{nl}\cdot 
\mathbf{G}^{\gtrless}_{ll}(E-\hbar\omega,k_z-q_z)\cdot
\nabla_j\mathbf{H}_{ln}\times\right.\nonumber\\
&&\phantom{i\sum_{q_z}\sum_{l}\sum_{ij}}
\left.\left(D^{\gtrless ij}_{ln}(\omega,q_z)-D^{\gtrless ij}_{ll}(\omega,q_z)
-D^{\gtrless ij}_{nn}(\omega,q_z)+D^{\gtrless
  ij}_{nl}(\omega,q_z)\right)\right].
\label{eq:3}
\end{eqnarray}
In Eq.~(\ref{eq:3}) $\nabla_i
\mathbf{H}_{nm}$=$\delta\mathbf{H}_{nm}/\delta(R_{m}^{i}-R_{n}^{i})$ is the
derivative of the Hamiltonian matrix block $\mathbf{H}_{nm}$ with
respect to variations along the $i$=$x$, $y$, or $z$ axis of the bond
$\mathbf{R}_{m}-\mathbf{R}_{n}$ connecting atoms $n$ and $m$. Since
the Hamiltonian matrix in the MLWF basis includes connections with the
30 (36) nearest-neighbors (NNs) of each atom in MoS$_2$ and WS$_2$ (black
phosphorus), the sum over $l$ in Eq.~(\ref{eq:3}) has to cover the
same range of interactions. Practically, summing over the 12 (MoS$_2$
and WS$_2$) or 13 (black phosphorus) NNs is sufficient. Still, although
the scattering self-energy $\mathbf{\Sigma}^{\gtrless S}(E,k_z)$ is
block diagonal only, its entries couple one atom at $\mathbf{R_i}$
with its 12-13 NNs at $\mathbf{R_l}$, thus indirectly accounting for
non-diagonal effects. The fact that each momentum $k_z$ is
connected to all possible $k_z-q_z$ points also contributes to the
presence of non-diagonal effects.

The phonon-electron scattering self-energy matrix
$\mathbf{\Pi}^{S}(\omega,q_z)$ cannot be assumed diagonal as
$\mathbf{\Sigma}^{\gtrless S}(E,k_z)$ because this would violate the
energy conservation rule between electrons and phonons. It has to take
the following form
\begin{eqnarray}
\Pi^{\gtrless ij}_{nn}(\omega,q_z)&=&-i\sum_{k_z}\sum_{l} \int\frac{dE}{2\pi} 
\mathrm{tr}\left\{\nabla_{i}\mathbf{H}_{ln}\cdot\mathbf{G}^{\gtrless}_{nn}(E+\hbar\omega,k_z+q_z)
\cdot\nabla_j\mathbf{H}_{nl}\cdot\mathbf{G}^{\lessgtr}_{ll}(E,k_z)\right.\nonumber\\
&&\phantom{-i\sum_{k_z}\sum_{l} \int}\left.+
\nabla_{i}\mathbf{H}_{nl}\cdot\mathbf{G}^{\gtrless}_{ll}(E+\hbar\omega,k_z+q_z)
\cdot\nabla_j\mathbf{H}_{ln}\cdot\mathbf{G}^{\lessgtr}_{nn}(E,k_z)\right\}
\label{eq:4}
\end{eqnarray}
for the diagonal block entries and
\begin{eqnarray}
\Pi^{\gtrless ij}_{nl}(\omega,q_z)&=&i\sum_{k_z}\int\frac{dE}{2\pi} 
\mathrm{tr}\left\{\nabla_{i}\mathbf{H}_{ln}\cdot\mathbf{G}^{\gtrless}_{nn}(E+\hbar\omega,k_z+q_z)
\cdot\nabla_j\mathbf{H}_{nl}\cdot\mathbf{G}^{\lessgtr}_{ll}(E,k_z)\right.\nonumber\\
&&\phantom{i\sum_{k_z}\int}\left.+
\nabla_{i}\mathbf{H}_{nl}\cdot\mathbf{G}^{\gtrless}_{ll}(E+\hbar\omega,k_z+q_z)
\cdot\nabla_j\mathbf{H}_{ln}\cdot\mathbf{G}^{\lessgtr}_{nn}(E,k_z)\right\}
\label{eq:5}
\end{eqnarray}
for the non-diagonal ones. In Eqs.~(\ref{eq:4}) and (\ref{eq:5}), ``$\mathrm{tr}\left\{\right\}$''
refers to the trace operator and atoms $l$ and $i$ must be coupled
with each other via a non-zero matrix element $\mathbf{H}_{li}$. The
last missing components are the retarded electron and phonon
scattering self-energies, which can be approximated as
\begin{eqnarray}
\mathbf{\Sigma}^{R}(E,k_z)&\approx&\frac{\left(\mathbf{\Sigma}^{>}(E,k_z)-
  \mathbf{\Sigma}^{<}(E,k_z)\right)}{2},\label{eq:6}\\
\mathbf{\Pi}^{R}(\omega,q_z)&\approx&\frac{\left(\mathbf{\Pi}^{>}(\omega,q_z)-
  \mathbf{\Pi}^{<}(\omega,q_z)\right)}{2}.\label{eq:7}
\end{eqnarray}
It is obvious from Eqs.~(\ref{eq:3}-\ref{eq:5}) that the scattering
self-energies bridge the electron and phonon populations because
$\mathbf{\Sigma}^{S}(E,k_z)$ depends on $\mathbf{D}(\omega,q_z)$ and
$\mathbf{\Pi}^{S}(\omega,q_z)$ on $\mathbf{G}(E,k_z)$. This is also
the reason why Eqs.~(\ref{eq:1}-\ref{eq:7}) must be self-consistently
solved till convergence is reached in the so-called Born
approximation. The resulting numerical problem is particularly
challenging since all energies $E$, frequencies $\omega$, and
momentum $k_z$/$q_z$ are connected altogether. Such systems of
equations cannot be tackled on a single computational core and must
therefore be treated in parallel. To manage the distribution of the
numerical tasks and the gathering of the needed data to calculate the
scattering self-energies, the multi-level parallelization scheme
introduced in Ref.~\cite{sc10} has been adapted and enhanced to enable
electro-thermal transport simulations of 2-D materials from
first-principles.

\subsection{Calculation of Observables}

A second self-consistent loop must be established between the solution
of the Schr\"odinger and Poisson equations to properly take the
electrostatics of the examined devices into account. This requires
evaluating the electron concentration $n(\mathbf{R}_i)$ (here only
$n$-type transistors are investigated) for each atomic position
$\mathbf{R}_i$ according to 
\begin{eqnarray}
n(\mathbf{R_i})&=&-i\sum_{k_z}\int \frac{dE}{2\pi}\mathrm{tr}
\left\{\mathbf{G}^{<}_{ii}(E,k_z)\right\}
\end{eqnarray}
and plugging the outcome into Poisson's equation expressed on a finite
element method (FEM) grid. After convergence the electrical current
flowing between two adjacent unit cells labeled $s$ and $s+1$ of the
2-D device structure 
\begin{eqnarray}
I_{d,s\rightarrow s+1}&=&\frac{q}{\hbar}\sum_{k_z}\sum_{i\in{s}}
\sum_{j\in{s+1}}\int\frac{dE}{2\pi}\phantom{.}\mathrm{tr}\left\{\mathbf{H}_{ij}\cdot\mathbf{G}^{<}_{ji}(E,k_z)-
\mathbf{G}^{<}_{ii}(E,k_z)\cdot\mathbf{H}_{ji}\right\},
\label{eq:9}
\end{eqnarray}
the electron component of the energy current between $s$ and $s+1$
\begin{eqnarray}
I_{E-el,s\rightarrow s+1}&=&\frac{1}{\hbar}\sum_{k_z}\sum_{i\in{s}}
\sum_{j\in{s+1}}\int\frac{dE}{2\pi}E\phantom{.}\mathrm{tr}\left\{\mathbf{H}_{ij}\cdot\mathbf{G}^{<}_{ji}(E,k_z)-
\mathbf{G}^{<}_{ii}(E,k_z)\cdot\mathbf{H}_{ji}\right\},
\label{eq:10}
\end{eqnarray}
as well as its phonon part
\begin{eqnarray}
I_{E-ph,s\rightarrow s+1}&=&\sum_{q_z}\sum_{i\in{s}}
\sum_{j\in{s+1}}\int\frac{d\omega}{2\pi}\hbar\omega\phantom{.}\mathrm{tr}\left\{\mathbf{\Phi}_{ij}
\cdot\mathbf{D}^{<}_{ji}(\omega,q_z)-\mathbf{D}^{<}_{ii}(\omega,q_z)\cdot\mathbf{\Phi}_{ji}\right\}
\label{eq:11}
\end{eqnarray}
can be computed from the electron and phonon Green's functions. In all
these equations, $\hbar$ is Planck's reduced constant, $q$ the
elementary charge, and it is implied that atom $i$ ($j$) belongs to
the unit cell $s$ ($s+1$). It should be pointed out in
Eqs.~(\ref{eq:9}) to (\ref{eq:11}) that (energy) current conservation
does not only require an integration over energy (or frequency), but
also a summation over the momentum ($k_z$ or $q_z$).

Another quantity of interest is the
effective lattice temperature of unit cell $s$, $T_{eff}(s)$. It can
be obtained by assuming that the total phonon population within one
unit cell $s$, $N_{ph}^{tot}(s)$, which is first calculated with NEGF,
locally obeys a Bose-Einstein distribution function
$N_{Bose}(T_{eff},\omega)=1/(\mathrm{exp}(\hbar\omega/k_BT_{eff})-1)$  
with $T_{eff}$ as the governing temperature and $k_B$ as Boltzmann's constant
\begin{eqnarray}
N_{ph}^{tot}(s)&=&\sum_{q_z}\sum_{i\in{s}}\int\frac{d(\hbar\omega)}{2\pi}
\mathrm{LDOS}(\mathbf{R}_i,\omega,q_z)N_{Bose}(T_{eff}(s),\omega)\label{eq:12}\\
&=&i\sum_{q_z}\sum_{i\in{s}}\int\frac{d(\hbar\omega)}{2\pi}
\mathrm{tr}\left\{\mathbf{D}^{<}_{ii}(\omega,q_z)\right\},\label{eq:13}
\end{eqnarray}
where the local density-of-states $\mathrm{LDOS}(\mathbf{R}_i,\omega,q_z)$ 
is defined as
\begin{eqnarray}
\mathrm{LDOS}(\mathbf{R}_i,\omega,q_z)&=&i\times\mathrm{tr}\left\{
\mathbf{D}^{>}_{ii}(\omega,q_z)-\mathbf{D}^{<}_{ii}(\omega,q_z)\right\}.
\end{eqnarray}
The effective lattice temperature of unit cell $s$ is then retrieved
by matching Eqs.~(\ref{eq:12}) and (\ref{eq:13}). The resulting
non-linear system of equations can be solved, for example, with a
Newton-Raphson iterative method. If necessary, an atomic resolution of
$T_{eff}$ is also possible.

\section{Device Results}\label{sec:results}

To illustrate the influence of electro-thermal effects on the
performance of $n$-type transistors made of a 2-D material, the
single-gate structure in Fig.~\ref{fig:fig1} has been simulated
with the models described in Section \ref{sec:method} together with
single-layer MoS$_2$, WS$_2$, and armchair-oriented black phosphorus
(AC BP) channels. The logic switch specifications include a gate
contact of length $L_g$=15 nm, 12.5 nm long source and drain
extensions with a donor doping concentration $N_D$=6e13 cm$^{-2}$,
perfectly ohmic contacts, a top (bottom) oxide layer of thickness
$t_{ox}$=3 (20) nm and relative permittivity $\epsilon_R$=20 (3.9),
and a supply voltage $V_{DD}$=0.67 V.

\begin{figure}
\includegraphics[width=0.45\linewidth]{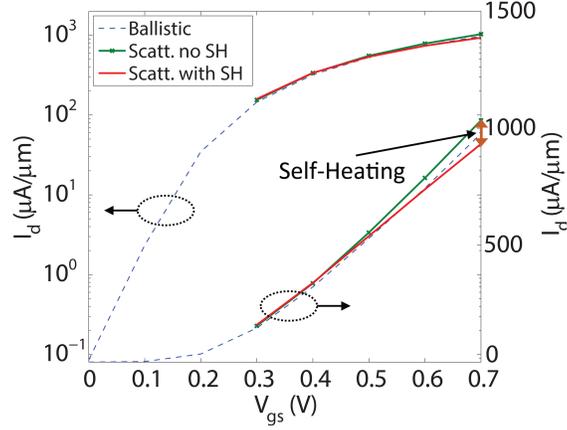}
\caption{\label{fig:fig4} (Color online) Transfer characteristics
  $I_d$-$V_{gs}$ at $V_{ds}$=0.67 V of the single-gate transistor in
  Fig.~\ref{fig:fig1} with a single-layer MoS$_2$ as channel. The
  ballistic limit of transport (dashed blue lines), the current with
  electron-phonon scattering, but an equilibrium phonon population
  (green lines with crosses), and the current with self-heating
  effects (solid red lines) are provided. The current reduction caused
  by self-heating is indicated by the orange double-arrow.}
\end{figure}

\begin{figure}
\includegraphics[width=0.45\linewidth]{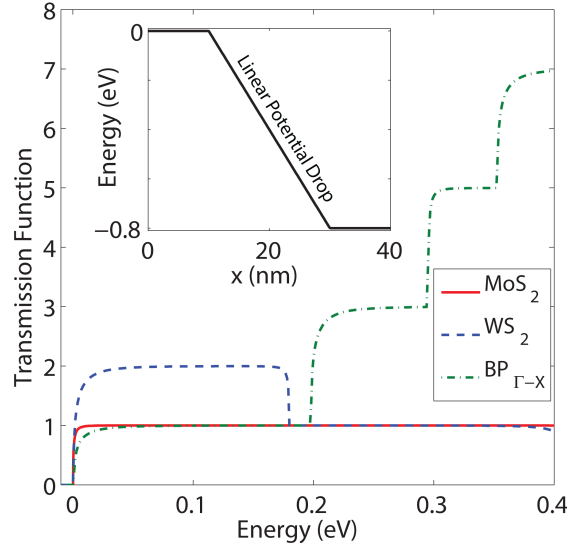}
\caption{\label{fig:fig5} (Color online) Electron transmission
  function, as obtained from quantum transport calculations, through
  the same single-layer 2-D materials as in Fig.~\ref{fig:fig3} at
  $k_z$=0, but with a linear potential drop (see inset) between both
  structure ends instead of flat band conditions.}
\end{figure}

The $I_d$-$V_{gs}$ transfer characteristics of the MoS$_2$ device at a
drain-to-source voltage $V_{ds}$=$V_{DD}$ and OFF-current
$I_{OFF}$=0.1 $\mu$A/$\mu$m are reported in Fig.~\ref{fig:fig4} for
three different situations: (i) ballistic transport
($\mathbf{\Sigma}^{S}(E,k_z)$=$\mathbf{\Pi}^{S}(\omega,q_z)$=0 in
Eqs.~(\ref{eq:1}-\ref{eq:4})), (ii) dissipative transport in the
presence of electron-phonon scattering, but with an equilibrium phonon
population (only $\mathbf{\Pi}^{S}(\omega,q_z)$=0), and (iii) in the
same configuration as in (ii), but with both the electron and phonon
populations driven out-of-equilibrium. The latter case incorporates
electro-thermal effects, contrary to the others where the lattice
temperature does not vary as a function of the electrical current and
remains equal to the room temperature $T_0$=300 K. It
might look surprising in Fig.~\ref{fig:fig4} that the ballistic
current is not the largest one. This peculiar feature has already
been discussed in details in Ref.~\cite{szabo1}: in typical transistor
structures, the ballistic transmission function $T(E)$ from left
(source) to right (drain) only slightly depends on the applied
$V_{ds}$ at a given $V_{gs}$. With strong drain-induced barrier
lowering (DIBL), a shift towards lower energy of the point where
$T(E)$ turns on is expected, but no shape changes. However, with
MoS$_2$ and most other TMDs, the transmission function exhibits a
strong dependence on $V_{ds}$, as can be seen by comparing $T(E)$ in
Fig.~\ref{fig:fig3} (flat-band conditions, zero electrostatic
potential everywhere) and in Fig.~\ref{fig:fig5}
(linear potential drop between both device ends). Obviously, at
energies close to the conduction band edge, one of the two available
channels in the source (sub-bands with a positive velocity) of MoS$_2$
stops conducting at high $V_{ds}$ so that $T(0)$=1 with a linear
potential drop, while $T(0)$=2 with flat bands. As a consequence, the
ballistic current is reduced. This ``anomaly'' is due to the
specificity of the TMD bandstructures, where sub-bands with a very
narrow energy width $\Delta E$ smaller than $qV_{ds}$ can be
found, as indicated in Fig.~\ref{fig:fig2}(a). These states are not
conductive in the ballistic limit of transport, but become active as
soon as electron-phonon scattering is switched on \cite{szabo1}. In
this case an electron occupying a non-conductive band can be
transferred to a conductive one by absorbing or emitting a
phonon. This explains the rather counterintuitive reinforcement of the
dissipative current with respect to the ballistic one in
Fig.~\ref{fig:fig4}.

\begin{figure}
\includegraphics[width=0.45\linewidth]{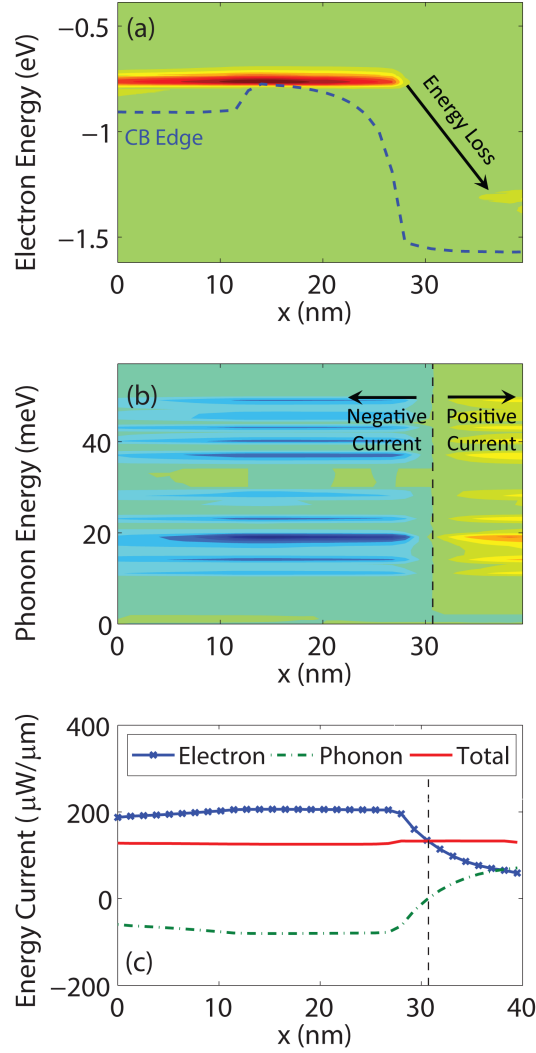}
\caption{\label{fig:fig6} (Color online) (a) Electron spectral current
  (current as a function of the position along the transport direction
  and of the electron energy) of the same transistor as in
  Fig.~\ref{fig:fig4} at $k_z$=0, $V_{gs}$=0.4 V, and $V_{ds}$=0.67
  V. The case with self-heating effects is plotted. Red indicates high
  current concentrations, green no current. The dashed blue line
  refers to the conduction band edge of MoS$_2$. Energy loss can be
  clearly seen on the drain (right) side of the device. (b) Phonon
  spectral current in the same structure as before. Red indicates now
  a positive current (from left to right), blue a negative one, green
  no current. The vertical dashed line corresponds to the location
  with no netto phonon current, i.e. the location with the highest
  phonon generation rate. (c)  Position-resolved electron (blue line
  with circles), phonon (dashed-dotted green line), and total
  (electron+phonon, solid red line) energy current flowing through the
  same transistor as before. The sign change of the phonon energy
  current at around $x$=31 nm as well as the conservation of the total
  energy current (almost constant solid red line) are clearly visible.}
\end{figure}

When the influence of the phonon-electron scattering self-energy
$\mathbf{\Pi}^{S}(\omega,q_z)$ is added to the picture, phonon
emission and absorption processes induce local variations
of the lattice temperature. This causes a current decrease by about
10\%, as compared to the case with electron-phonon interactions
only, which is marked by a double arrow in Fig.~\ref{fig:fig4} and
labeled self-heating. To better understand what happens inside the
MoS$_2$ transistor, the spectral electron and phonon current
distributions in Fig.~\ref{fig:fig6}(a-b) are investigated as next
step. It can be observed in sub-plot (a) that electrons loose a
substantial portion of their energy through phonon emission after
passing the maximum of the electrostatic potential energy barrier, when they
are accelerated by the strong electric field on the drain side. From
sub-plot (b) it can be inferred that at the location with the highest
phonon generation rate (and therefore the highest lattice temperature
too, as explained later) at around $x$=31 nm, the phonon current is
equal to 0 and the created lattice vibrations either propagate towards
the drain ($x$=40 nm, positive current) or the source ($x$=0 nm,
negative current) contact. The phonons that reach the source can
interact there with electrons moving towards the drain and increase
their backscattering rate 
\cite{prb14}, thus leading to the current reduction visible in
Fig.~\ref{fig:fig4}. To complement this analysis, it is demonstrated
in Fig.~\ref{fig:fig6}(c) that energy is conserved in the developed
simulation approach. The sum of the electron and phonon energy
currents, as given in Eqs.~(\ref{eq:10}) and (\ref{eq:11}), 
is indeed constant from source to drain. This indicates that the
energy lost by electrons is correctly transferred to the phonon bath
and vice versa. Slight fluctuations of the total energy
current still exist (less than 1\%): they are due to the
slow convergence of Eqs.~(\ref{eq:1}-\ref{eq:6}). It has been verified
that more self-consistent iterations between the Green's Functions and the
scattering self-energies improve the situation, at the expense
of the computational time.

\begin{figure}
\includegraphics[width=0.45\linewidth]{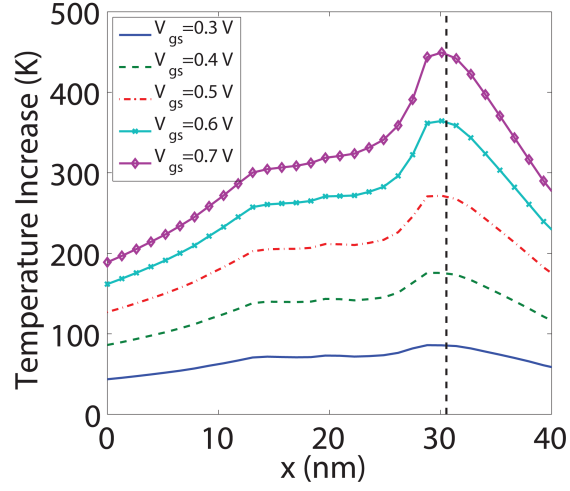}
\caption{\label{fig:fig7} (Color online) Effective lattice temperature
  increase $\Delta T$=$T$-$T_0$ as a function of the position along
  the $x$-axis and of the gate voltage $V_{gs}$ at a fixed drain bias
  $V_{ds}$=0.67 V for the same single-layer MoS$_2$ transistor as in
  Fig.~\ref{fig:fig6}.}
\end{figure}

With Eqs.~(\ref{eq:12}) and (\ref{eq:13}), the non-equilibrium phonon
population can be converted into a position-resolved effective lattice
temperature $T_{eff}$ that is resolved at the unit cell level. The
results are provided in Fig.~\ref{fig:fig7} for the MoS$_2$
transistor. What is shown is the lattice temperature increase $\Delta
T$=$T_{eff}$-$T_0$ as a function of the location along the $x$-axis and
as a function of the applied gate-to-source voltage $V_{gs}$. Two
points should be emphasized: first, $T_{eff}$ peaks at the same
$x$-coordinate as where the phonon current vanishes in
Fig.~\ref{fig:fig6}(b) and where the phonon generation rate reaches
its maximum. The formed hot spot is in fact situated close to the
channel-drain interface, in the region with the sharpest drop of the
electrostatic potential energy and therefore the highest phonon
emission probability. Secondly, the $\Delta T$ values are extremely
high $>$400 K at $V_{gs}$=0.7 V. Before approaching such temperatures,
the MoS$_2$ single-layer crystal would already have oxidized (oxidation
temperature: 675 K) \cite{zhou} and the whole device would have broken
down. As a comparison, in Ref.~\cite{pop2}, the $\Delta T$ of a
micrometer-scale monolayer MoS$_2$ transistor has been estimated through
Raman thermometry measurements to be around 250 K for a much larger
$V_{ds}$=30 V vs. 0.67 V here, but a smaller current $I_d\approx$200
$\mu$A/$\mu$m vs. $\sim$1000 $\mu$A/$\mu$m here and a higher contact
resistance $R_c$=5k$\Omega\times\mu$m vs. perfectly ohmic contacts
here. This confirms that self-heating effects are very important in
monolayer MoS$_2$, but that the simulated temperature increases are
overestimated, mainly because no  phonon escape through the
surrounding oxide layers is allowed. Such mechanisms would
definitively lead to a decrease of the effective lattice temperature,
but their exact contribution is difficult to quantify.

\begin{figure}
\includegraphics[width=\linewidth]{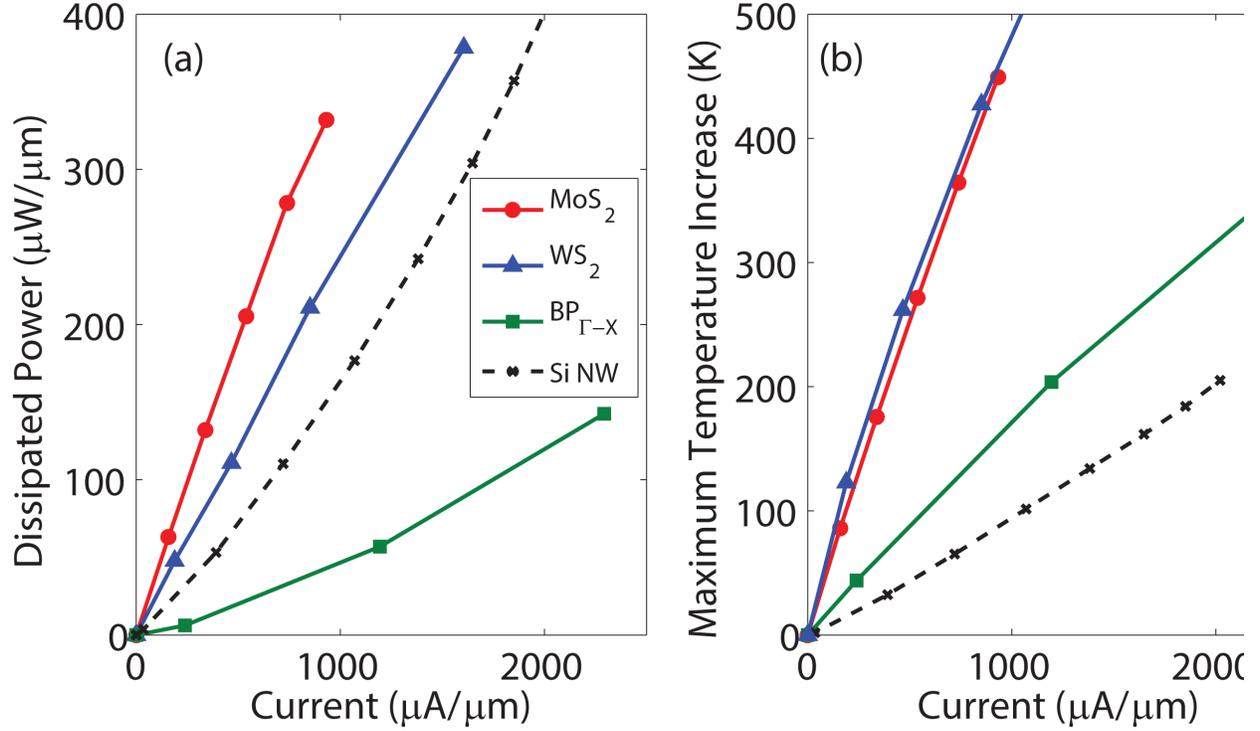}
\caption{\label{fig:fig8} (Color online) (a) Power dissipated inside
  the channel of a single-layer MoS$_2$ (red line with circles),
  WS$_2$ (blue line with triangles), and AC BP (green line with
  squares) transistor with the same structure and dimensions as in
  Fig.~\ref{fig:fig1} vs. the electrical current flowing through this
  device. The data for a circular Si nanowire transistor with a
  gate-all-around configuration, a diameter $d$=3 nm, and transport
  along the $<$100$>$ crystal axis are also provided (dashed black
  line) \cite{prb14}. (b) Same as (a), but for the maximum lattice
  temperature increase vs. drive current. (c) Combination of (a) and (b),
  i.e. maximum lattice temperature increase vs. dissipated power
  inside the channel.}
\end{figure}

Keeping in mind that $\Delta T$ is overestimated, it is still
relevant to compare the electro-thermal properties of MoS$_2$ with
those of other 2-D materials, namely WS$_2$ and AC black phosphorus,
and with those of more conventional ultra-scaled Si nanowire transistors
since the same set of approximations is applied everywhere,
i.e. phonons escape at the source and drain only, not through oxide
layers. The goal of this study is to determine what material/design is
the least sensitive to self-heating and why. In Fig.~\ref{fig:fig8}(a)
the electrical power dissipated inside the channel of all switches,
$P_{diss}$, is plotted as a function of the drive current $I_d$. For
the Si nanowire transistor, a $<$100$>$-oriented structure with a
diameter $d$=3nm, a gate length $L_g$=15 nm, and a gate-all-around
configuration has been selected and the data from Ref.~\cite{prb14}
have been recalled. Here, $P_{diss}$ is defined as the difference in the
electron energy current (see Eq.~(\ref{eq:10})) between the source and 
the drain. Normally, it is expected that the total dissipated power is
equal to $V_{ds}\times I_d$ and that it has the same value for all structures
at a given $I_d$. However, due to the short length of the considered
devices, only a fraction of the total power dissipation takes place
in the channel region, the rest in the contacts. This explains why
$P_{diss}<V_{ds}\times I_d$ in our simulations and why the curves are
different in Fig.~\ref{fig:fig8}(a).

\begin{figure}
\includegraphics[width=0.45\linewidth]{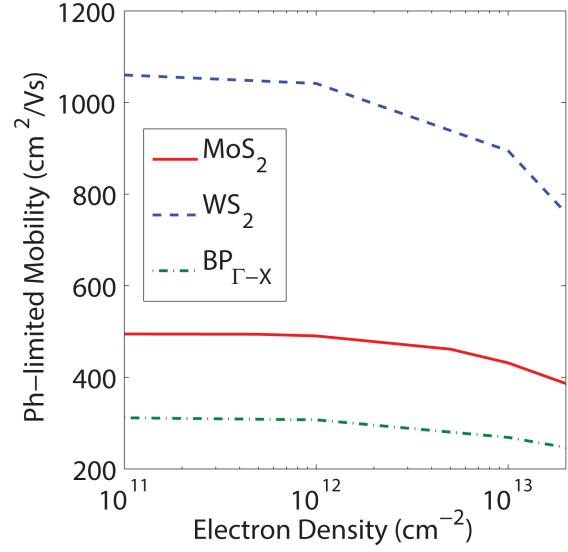}
\caption{\label{fig:fig9} (Color online) Electron phonon-limited
  mobility of single-layer MoS$_2$, WS$_2$, and AC BP as a function of
  the carrier concentration \cite{iedm16} and with the same plotting
  conventions as in Fig.~\ref{fig:fig3}. The ``dR/dL'' mobility
  extraction method has been used for that purpose \cite{rim}.}
\end{figure}

The amount of dissipated power in a given component can be put in
relation with the phonon-limited mobility $\mu_{ph}$ of its underlying
channel material \cite{nl16}: the higher $\mu_{ph}$ the less phonons are
emitted and the less electrical power is converted into heat. To
verify whether this statement is valid or not here, the phonon-limited
electron mobility of single-layer MoS$_2$, WS$_2$, and AC black
phosphorus, as computed with the ``dR/dL'' method \cite{rim}, is
presented in Fig.~\ref{fig:fig9} as a function of the
carrier concentration. Due to the heavier atomic mass of tungsten as
compared to molybdenum, the amplitude of the crystal vibrations and
in turn the electron-phonon coupling strength is weaker in WS$_2$
than in MoS$_2$, which gives a higher mobility and a lower dissipated
power at a fixed current magnitude, as postulated above. This
demonstration fails however when AC black phosphorus comes into play:
its mobility is smaller than that of MoS$_2$ and WS$_2$ due to more
significant crystal vibrations, but it still dissipates less
power. The reason behind this apparent inconsistency can be traced
back to the distinctive bandstructure of TMDs, where sub-bands with a
narrow energy width cannot carry current at high
$V_{ds}$, as shown in Fig.~\ref{fig:fig5}. As a result,
electron-phonon interactions are needed to make those sub-bands
conductive and to reach high current densities. The number of phonon
emission processes occurring in TMDs under the application of a high
$V_{ds}$ is therefore much more important than predicted by the
mobility, which is computed under flat-band conditions. While
electron-phonon interactions contribute to a current increase in TMDs,
they simultaneously consume electrical power and lead to the higher heat
dissipation of MoS$_2$ and WS$_2$ seen in Fig.~\ref{fig:fig8}(a). 
Although the Si nanowire transistor has about the same phonon-limited
electron mobility as AC black phosphorus (300 cm$^2$/Vs at an electron
concentration $n_D\approx$1e13 cm$^{-2}$), it dissipates more
power. Since the crystal and device structures are quite different, it
is not possible to explain this behavior with the same line of
arguments as before.

Figure \ref{fig:fig8}(b) then reports the maximum effective lattice
temperature increase extracted from the same transistors as before
versus the electrical current $I_d$. What matters in this sub-plot is
no more the number of phonons that is generated (heat dissipation),
but once they have been created how efficiently they can leave the active
region of the logic switch. If the emitted phonons have a high group
velocity, they more rapidly diffuse away from the region where they
originate such that the effective lattice temperature at this location
cannot rise as much as it would have if the phonon population would keep
accumulating. Hence, the thermal transport properties determine to a
large extend the magnitude of the self-heating effects. For example,
in WS$_2$, less power is dissipated at a given $I_d$ than in MoS$_2$,
but the lattice temperature increase $\Delta T$ is the same in both
2-D materials: less crystal vibrations are generated in WS$_2$ due to
the weaker electron-phonon coupling and lower electron effective mass,
but they propagate slower due to the higher atomic mass of
tungsten. The poorer thermal transport properties of single-layer
WS$_2$ are indirectly illustrated in Table \ref{tab:tab1}, which
gives the sound velocity of all studied 2-D crystals, as derived
from \textit{ab-initio} calculations. It can also be
seen in Fig.~\ref{fig:fig8}(b) that the temperature increase is lower
in AC black phosphorus and in the Si nanowire than in both TMDs.

Finally, the maximum lattice temperature increase can be represented
as a function of the power dissipated inside the transistor
channel. The results are shown in Fig.~\ref{fig:fig8}(c). While the
curves for the 2-D materials are not identical, they all exhibit a
much higher lattice temperature increase than the Si nanowire at a
given $P_{diss}$. The strong confinement of phonons in 2-D crystals
leads to rather poor thermal transport properties and prevents the
rapid evacuation of heat from the active region of 2-D logic
switches. As mentioned earlier, allowing phonons to escape through the
surrounding oxide layers would certainly reduce the $\Delta T$, but
the same would happen in the Si nanowire where phonons are also
confined in the semiconductor \cite{apl17}. A recent experimental
study on single-layer MoS$_2$ \cite{pop2} has clearly established
that self-heating is much more significant in this 2-D material than
in Si, which qualitatively agrees with the finding of this paper.

\section{Conclusion}\label{sec:conc}

An \textit{ab-initio} electro-thermal transport modeling approach has
been introduced in this paper and applied to the simulation of
single-gate transistors with a single-layer 2-D material as
channel. After demonstrating the functionality of the implemented
electron and phonon quantum transport solver, the effective lattice
temperature and the power dissipation of logic switches made of
MoS$_2$, WS$_2$, AC black phosphorus monolayers as well as of a Si
nanowire have been computed and compared to each other. It turns out
that the considered 2-D semiconductors are much more affected by
self-heating than Silicon, especially at high current
densities. This might become a serious problem in the future when the
contact series resistances of devices with a 2-D channel material will
shrink and their drive current increase. Enhancing heat dissipation
through the bottom oxide layer could be the solution to mitigate this issue,
for example by carefully chosen the oxide material or by making it
thinner. Further model developments will be needed to account for
these effects and support the on-going experimental activity.

\section*{Acknowledgment}
This work was supported by the Swiss National Science Foundation Grant
No.~PP00P2\_159314, by ETH Research Grant ETH-32 15-1, by the
European Research Council under Grant Agreement No 335684-E-MOBILE, and by CSCS
under Projects s662 and s714.

\clearpage

\clearpage

\clearpage

\clearpage

\clearpage

\clearpage

\clearpage

\clearpage

\clearpage

\clearpage

\begin{table}
\begin{tabular}{|c|c|c|c|c|c|c|}
\hline
&$E_g$ (eV)&$m_{e}^{*}$&$m_{h}^{*}$&$v_{s,TA}$ (km/s)&$v_{s,LA}$
(km/s)&$\hbar \omega_{opt}$ (meV)\\
\hline
MoS$_2$&1.67&0.46&0.56&4.7&7.1&58\\
\hline
WS$_2$&1.8&0.3&0.38&3.9&6.2&53\\
\hline
BP$_{\Gamma-X}$&1.61&0.18&0.15&4.6&5.0&57\\
\hline
BP$_{\Gamma-Y}$&1.61&1.2&2.5&3.9&8.5&57\\
\hline
\end{tabular}
\caption{\label{tab:tab1} Selected material parameters of single-layer
  MoS$_2$, WS$_2$, and black phosphorus along the $\Gamma$-$X$ and
  $\Gamma$-$Y$ symmetry lines: band gap ($E_g$), electron ($m_e$) and 
  hole ($m_h$) effective masses at the band edges, sound velocity of
  the transverse ($v_{s,TA}$) and longitudinal ($v_{s,LA}$) acoustic
  phonon branches, as well as optical phonon energy ($\hbar
  \omega_{opt}$). All data were extracted from DFT \cite{vasp} and
  DFPT \cite{dfpt} calculations.}
\end{table}

\clearpage


\begin{thebibliography}{1}
\bibitem{mounet}
N.~Mounet, M.~Gibertini, P.~Schwaller, A.~Merkys, I.~E.~Castelli,
A.~Cepellotti, G.~Pizzi, and N.~Marzari, arXiv:1611.05234 (2016).
 \bibitem{kuhn}
K.~J.~Kuhn, IEEE Trans.~Elec.~Dev. 59, 1813 (2012).
\bibitem{novoselov}
K.~S.~Novoselov, A.~K.~Geim, S.~V.~Morozov, D.~Jiang, Y.~Zhang,
S.~V.~Dubonos, I.~V.~Grigorieva, and A.~A.~Firsov, Science 306,
666-669 (2004).
\bibitem{kis}
B.~Radisavljevic, A.~Radenovic, J.~Brivio, V.~Giacometti, and A.~Kis,
Nature Nanotech. 6, 147-150 (2011).
\bibitem{iqbal}
M.~W.~Iqbal, M.~Z.~Iqbal, M.~F.~Khan, M.~A.~Shehzad, Y.~Seo,
J.~H.~Park, C.~Hwang, and J.~Eom, Sci.~Rep. 5, 10699 (2015). 
\bibitem{miro}
P.~Mir\'o, M.~Audiffred, and T.~Heine, Chemical Society Reviews 43,
6537-6554 (2014).
\bibitem{pop0}
E.~Pop, Nano Res 3, 147-169 (2010).
\bibitem{balandin}
A.~A.~Balandin, S.~Ghosh, W.~Bao, I.~Calizo, D.~Teweldebrhan, F.~Miao,
and C.~N.~Lau, Nano Letters 8, 902-907 (2008).
\bibitem{shi}
J.~H.~Seol, I.~Jo, A.~L.~Moore, L.~Lindsay, Z.~H.~Aitken,
M.~T.~Pettes, X.~Li, Z.~Yao, R.~Huang, D.~Broido, N.~Mingo,
R.~S.~Ruoff, and L.~Shi, Science 328, 213-216 (2010).
\bibitem{pop1}
E.~Pop, V.~Varshney, A.~K.~Roy, MRS Bulletin 37, 1273 (2012).
\bibitem{xing}
R.~Yan, J.~R.~Simpson, S.~Bertolazzi, J.~Brivio⊥, M.~Watson, X.~Wu,
A.~Kis, T.~Luo, A.~R.~H.~Walker, and H.~G.~Xing, ACS Nano 8, 986-993 
(2014).
\bibitem{pop2}
E.~Yalon, C.~J.~McClellan, K.~K.~H.~Smithe, M.~Mu\~noz Rojo, R.~Lily
Xu, S.~V.~Suryavanshi, A.~J.~Gabourie, C.~M.~Neumann, F.~Xiong,
A.~Barati Farimani, and E.~Pop,  Nano Lett. 17, 3429 (2017).
\bibitem{cahill}
H.~Jang, J.~D.~Wood, C.~R.~Ryder, M.~C.~Hersam, and D.~G.~Cahill,
Adv. Materials 27, 8017 (2015).
\bibitem{datta}
S.~Datta, ``Electronic Transport in Mesoscopic Systems'', Cambridge
Press (1995).
\bibitem{hguo}
F.~Zahid, L.~Liu, Y.~Zhu, J.~Wang, and H.~Guo, AIP Advances 3, 052111
(2013). 
\bibitem{klimeck}
H.~Ilatikhameneh, Y.~Tan, B.~Novakovic, G.~Klimeck, R.~Rahman, and
J.~Appenzeller, IEEE Journal on Exploratory Solid-State Computational
Devices and Circuits 1, 12 (2015).
\bibitem{register}
J.~Chang, L.~F.~Register, and S.~K.~Banerjee, J.~Appl.~Phys. 115,
084506 (2014).  
\bibitem{fiori1}
G.~Pizzi, M.~Gibertini, E.~Dib, N.~Marzari, G.~Iannaccone, and
G.~Fiori, Nature Comm. 7, 12585 (2016).
\bibitem{qw0}
W.~S.~Leong, X.~Luo, Y.~Li, K.~H.~Khoo, S.~Y.~Quek and J.~T.~L.~Thong,
ACS Nano 9, 869 (2014).  
\bibitem{gnani}
E.~Gnani, A.~Gnudi, S.~Reggiani, M.~Luisier, and G.~Baccarani,
IEEE Trans.~Nanotech 7, 700 (2008).
\bibitem{szabo1}
A.~Szabo, R.~Rhyner, and M.~Luisier, Phys.~Rev.~B 92, 035435 (2015).
\bibitem{kaas}
K.~Kaasbjerg, K.~Thygesen, and K.~W.~Jacobsen, Phys.~Rev.~B 85, 115317
(2012).
\bibitem{qw}
T.~Gunst, T.~Markussen, K.~Stokbro, and M.~Brandbyge, Phys.~Rev.~B
93, 035414 (2016).
\bibitem{marzari}
G.~Pizzi, D.~Volja, B.~Kozinsky, M.~Fornari, and N.~Marzari, Computer
Physics Communications 185, 422-429 (2014). 
\bibitem{fiori2}
G.~Fiori, F.~Bonaccorso, G.~Iannaccone, T.~Palacios, D.~Neumaier,
A.~Seabaugh, S.~K.~Banerjee, and L.~Colombo, Nature Nano. 9, 768-779
(2014). 
\bibitem{md1}
Y.~Hong, J.~Zhang, and X.~C.~Zeng, J.~Phys.~Chem.~C 120 26067 (2016).
\bibitem{md2}
X.~Liu, G.~Zhanga, Q.-X.~Pei, and Y.-W.~Zhang, Appl.~Phys.~Lett. 103,
133113 (2013). 
\bibitem{lbt1}
B.~Peng, H.~Zhang, H.~Shao, Y.~Xu, X.~Zhang, and H.~Zhu, RSC Adv. 6, 5767
(2016).
\bibitem{lbt2}
W.~Li, J.~Carrete, and N.~Mingo, Appl.~Phys.~Lett. 103, 253103 (2013).
\bibitem{marzari2}
A.~Cepellotti, G.~Fugallo, L.~Paulatto, M.~Lazzeri, F.~Mauri, and
N.~Marzari, Nature Comm. 6, 6400 (2015).
\bibitem{asai}
Y.~Asai, Phys.~Rev.~B 78, 045434 (2008).
\bibitem{pecchia}
A.~Pecchia, G.~Romano, and A.~Di Carlo, Phys.~Rev.~B 75, 035401
(2007).
\bibitem{frederiksen}
T.~Frederiksen, M.~Paulsson, M.~Brandbyge, and A.-P.~Jauho,
Phys.~Rev.~B 75, 205413 (2007).
\bibitem{jtlu1}
J.~T.~L\"u, and J.-S.~Wang, Phys.~Rev.~B 76, 165418 (2007).
\bibitem{jtlu2}
J.-T.L\"u, R.~B.~Christensen, J.-S.~Wang, P.~Hedegard, and
M.~Brandbyge, Phys.~Rev.~Lett. 114, 096801 (2015).
\bibitem{prb14}
R.~Rhyner and M.~Luisier, Phys,~Rev.~B 89, 235311 (2014).
\bibitem{vasp}
G.~Kresse and J.~Furthm\"uller, Comput.~Mat.~Sci. 6, 15 (1996).
\bibitem{wannier90}
A.~A.~Mostofi, J.~R.~Yates, Y.-S.~Lee, I.~Souza, D.~Vanderbilt, and
N.~Marzari, Comput.~Phys.~Commun. 178, 685 (2008).
\bibitem{pbe}
J.~P.~Perdew, K.~Burke, and M.~Ernzerhof, Phys.~Rev.~Lett. 77 3865
(1996).
\bibitem{hse06}
J.~Heyd, G.~E.~Scuseria, and M.~Ernzerhof,  J.~Chem.~Phys. 124, 219906
(2006).
\bibitem{dfpt}
A.~Togo, F.~Oba, and I.~Tanaka, Phys.~Rev.~B 78, 134106 (2008).
\bibitem{rgf}
A.~Svizhenko, M.~P.~Anantram, T.~R.~Govindan, B.~Biegel, and
R.~Venugopal, J.~App.~Phys. 91, 2343-2354 (2002).
\bibitem{prb06}
M.~Luisier, A.~Schenk, W.~Fichtner, and G.~Klimeck, Phys.~Rev.~B 74,
205323 (2006).
\bibitem{apl17}
R.~Rhyner and M.~Luisier, Appl.~Phys.~Lett. 110, 103508 (2017).
\bibitem{sc10}
M.~Luisier, SC '10 Proceedings of the 2010 ACM/IEEE International
Conference for High Performance Computing, Networking, Storage and
Analysis Pages 1-11 (2010).
\bibitem{zhou}
H. Zhou, F.!Yu, Y.~Liu, X.~Zou, C.~Cong, C.~Qiu, T.~Yu, Z.~Yan,
X.~Shen, L.~Sun, B.~I.~Yakobson, and J.~M.~Tour, Nano Res. 6, 703
(2013).
\bibitem{nl16}
R.~Rhyner and M. Luisier, Nano Letters 16, 1022-1026 (2016).
\bibitem{rim}
K.~Rim, S.~Narasimha, M.~Longstreet, A.~Mocuta, and J.~Cai,
Proceedings of the IEDM Conference 2002, pp.~43-46 (2002).
\bibitem{iedm16}
M.~Luisier, A.~Szabo, C.~Stieger, C.~Klinkert, S.~Br\"uck, A.~Jain,
and L.~Novotny, Proceedings of the IEDM Conference 2016,
pp.~5.4.1-5.4.4 (2016).
\end{thebibliography}
\end{document}